\documentclass{article}
\usepackage{spconf,amsmath,graphicx}

\usepackage{amssymb,amsfonts,amsmath}
\usepackage{mathtools}
\usepackage[linesnumbered,ruled,vlined]{algorithm2e}
\usepackage{arydshln}
\usepackage{booktabs}
\usepackage[bookmarks=false]{hyperref}
\usepackage{tikz}
\usetikzlibrary{fit,calc,positioning,decorations.pathreplacing,matrix}
\usetikzlibrary{shapes,arrows}
\usetikzlibrary{calc}
\usepackage{multirow}
\usepackage{balance}

\hypersetup{
    colorlinks,
    linkcolor={black},
    citecolor={black},
    urlcolor={black},
}

\newcommand{\w}{\mathbf{w}}
\newcommand{\smlv}{\mathbf{v}}
\newcommand{\n}{\mathbf{n}}
\newcommand{\x}{\mathbf{x}}
\newcommand{\smlu}{\mathbf{u}}
\newcommand{\s}{\mathbf{s}}
\newcommand{\e}{\mathbf{e}}
\newcommand{\y}{\mathbf{y}}

\newcommand{\kk}{\mathbf{k}}
\newcommand{\q}{\mathbf{q}}

\newcommand{\Bign}{\mathbf{N}}
\newcommand{\Bigs}{\mathbf{S}}
\newcommand{\Bigmm}{\mathbf{M}}
\newcommand{\Bigy}{\mathbf{Y}}
\newcommand{\Bigp}{\mathbf{P}}
\newcommand{\Bigu}{\mathbf{U}}

\newcommand{\R}{{\mathbb{R}}}

\newcommand{\ww}{\mathbf{W}}

\usepackage{multirow}
\usepackage{lipsum}

\newcommand\blfootnote[1]{%
	\begingroup
	\renewcommand\thefootnote{}\footnote{#1}%
	\addtocounter{footnote}{-1}%
	\endgroup
}

\title{Channel-Attention Dense U-Net for Multichannel Speech Enhancement}

\name{Bahareh Tolooshams$^{1}$, Ritwik Giri$^{2}$, Andrew H. Song$^{3}$, Umut Isik$^{2}$, and Arvindh Krishnaswamy$^{2}$}

\address{$^1$School of Engineering and Applied Sciences, Harvard University, Cambridge, MA \\$^2$Amazon Web Services, Palo Alto, CA\\$^3$Massachusetts Institute of Technology, Cambridge, MA}
\begin{document}

\sloppy

\setlength{\abovedisplayskip}{1mm}
\setlength{\belowdisplayskip}{1mm}

\ninept
\maketitle
\begin{abstract}

Supervised deep learning has gained significant attention for speech enhancement recently. The state-of-the-art deep learning methods perform the task by learning a ratio/binary mask that is applied to the mixture in the time-frequency domain to produce the clean speech. Despite the great performance in the single-channel setting, these frameworks lag in performance in the multichannel setting as the majority of these methods a) fail to exploit the available spatial information fully, and b) still treat the deep architecture as a black box which may not be well-suited for multichannel audio processing. This paper addresses these drawbacks, a) by utilizing complex ratio masking instead of masking on the magnitude of the spectrogram, and more importantly, b) by introducing a channel-attention mechanism inside the deep architecture to mimic beamforming. We propose Channel-Attention Dense U-Net, in which we apply the channel-attention unit recursively on feature maps at every layer of the network, enabling the network to perform \emph{non-linear} beamforming. We demonstrate the superior performance of the network against the state-of-the-art approaches on the CHiME-3 dataset.

\end{abstract}

\blfootnote{This work was done while B. Tolooshams and A. H. Song were interns at Amazon Web Services.}

\section{Introduction}
\label{sec:intro}


Multichannel speech enhancement is the problem of obtaining a clean speech estimate from  multiple channels of noisy mixture recordings. Traditionally, beamforming techniques have been employed, where a linear spatial filter is estimated, per frequency, to boost the signal from the desired target direction while attenuating the interferences from other directions by utilizing second-order statistics, e.g., spatial covariance of speech and noise~\cite{benesty2008microphone}.


In recent years, deep learning (DL) based supervised speech enhancement techniques have achieved significant success~\cite{wang2018supervised}, specifically for monaural/single-channel case. Motivated by this success, a recent line of work proposes to combine supervised single-channel techniques with unsupervised beamforming methods for multichannel case~\cite{erdogan, heymann2016neural}. These approaches are broadly known as neural beamforming, where a neural network estimates the second-order statistics of speech and noise, using estimated time-frequency (TF) masks, after which the beamformer is applied to linearly combine the multichannel mixture to produce clean speech. 
However, the performance of neural beamforming is limited by the nature of beamforming, a \textit{linear} spatial filter per frequency bin.

Another line of work \cite{wang2018combining, wang2018multi} proposes to use spatial features along with spectral information to estimate TF masks. Most of these approaches have an explicit step to extract spatial features such as interchannel time/phase/level difference (ITD/IPD/ILD). Recent work~\cite{endtoend} automatically extracts phase information from the input mixture by incorporating IPD as a block inside the neural network. \cite{maskchak} takes a more general approach to predict the TF mask by directly feeding magnitude and phase of the complex spectrogram from all microphones to a convolutional neural network (CNN). Despite incorporating spatial information, these methods still focus on predicting a real mask, hence resort to using the noisy phase, and ignore phase-enhancement.

To overcome the aforementioned limitations, this paper proposes an end-to-end neural architecture for multichannel speech enhancement, which we call Channel-Attention Dense U-Net. The distinguishing feature of the proposed framework is a Channel-Attention (CA) mechanism inspired by beamforming. CA is motivated by the self-attention mechanism, which captures global dependencies within the data. Self-attention has been previously used in various fields~\cite{bahdanau2014neural, parikhetal2016decomposable,saGANs}, as well as speech enhancement in the single-channel setting~\cite{attentionunet}. This paper incorporates CA into a CNN to guide the network to decide, at every layer, which feature maps to pay the most attention to. This work, therefore, extends the idea of beamforming on the input space to a latent space. 

In addition to the CA units, the network is a variation of U-Net~\cite{unet}, a popular architecture for source separation, and DenseNet~\cite{denseNet}. Motivated by the success of complex ratio masking in the single-channel case \cite{crm}, our approach takes both real and imaginary part of the complex mixture short-time Fourier transform (STFT) and estimates a complex ratio mask (CRM) unlike in~\cite{wang2018multi, TFmask}. The CRM is then applied to the mixture STFT to obtain the clean speech. Channel-Attention Dense U-Net does not require an explicit spatial feature extraction step; instead it implicitly identifies and exploits the relevant spatial information.

Rest of the paper is organized as follows: Section~\ref{sec:saunet} introduces the proposed network, Channel-Attention Dense U-Net, and discusses mechanism of CA in detail. Section~\ref{sec:exp} describes the dataset, network parameters, and evaluation criteria. This is followed by Section~\ref{sec:res}, in which we demonstrate the outperformance of our network against state-of-the-art methods. Finally, Section~\ref{sec:con} concludes the paper and discusses some future directions of this work.



\section{Channel-Attention Dense U-Net}
\label{sec:saunet}
\subsection{Problem Description}
\label{sec:model}
Let $\y^c \in \R^{N}$ be the discrete-time signal of a noisy mixture at microphone $c$. We assume that $\{\y^c\}_{c=1}^C$, for $c=1,\ldots,C$ and $n=1,\ldots,N$, follows the generative model
\begin{equation}\label{eq:gen}
\y^{c}[n] = \s^{c}[n] + \n^{c}[n],
\end{equation}
\noindent where $\s^{c}[n]$ and $\n^{c}[n]$ represent the clean speech and noise recorded at channel $c$, at time $n$, respectively. The goal of speech enhancement is to estimate $\hat{\s}^{\text{ref}}$, where $\text{ref}\in\{1,\ldots,C\}$ denotes a reference channel from the multichannel mixtures $\{\y^c\}_{c=1}^C$. We also denote $\Bigy^c\in \mathbb{C}^{F\times T}$ as the STFT of $\y^c$, where $F$ and $T$ are the number of frequency bins and time frames, respectively, and $\Bigy = [\Bigy^1, \ldots, \Bigy^C]  \in \mathbb{C}^{F \times T \times C}$ as multichannel STFT.

Let $\Bigy_f = [\Bigy^1_f, \ldots, \Bigy^C_f]\in \mathbb{C}^{T \times C}$ be the multichannel STFT at frequency bin $f$, the traditional beamformers, such as the popular MVDR beamformer, \emph{linearly} combine $\{\Bigy_f^c\}_{c=1}^C$ with the estimated beamforming weights $\hat{\w}_f\in\mathbb{C}^C$, to produce the estimated clean speech $\hat{\Bigs}_f$ at each frequency $f$ (e.i., $\hat{\Bigs}_f=\Bigy_f\hat{\w}_f^{\text{H}}\in\mathbb{C}^{T}$). As will be made clearer in subsequent sections, our proposed framework applies attention weights, i.e., a weight matrix similar to beamforming weights, recursively to the multichannel input and feature maps, extending the beamforming analogy to a \emph{non-linear} combination.
\vspace{-2mm}
\subsection{Framework Overview}
\label{sec:arc}
Channel-Attention Dense U-Net consists of an encoder, a mask estimation network, and a decoder. The encoder performs STFT on the mixture $\{\y^c\}_{c=1}^C$ to produce $\Bigy$. Given $\Bigy$, the mask estimation network computes both the speech mask $\mathbf{M}$ and noise mask $\mathbf{M}^{\text{noise}}$, which are multiplied to input, to obtain the clean speech estimate $\hat \Bigs \in \mathbb{C}^{F \times T \times C}$ and the noise estimate $\hat \Bign \in \mathbb{C}^{F \times T \times C}$. Finally, the decoder performs inverse-STFT on $\hat \Bigs$ and $\hat \Bign$ to produce time-domain estimates of the speech $\hat \s \in \R^{N \times C}$ and the noise $\hat \n \in \R^{N \times C}$.



We now expand on how the outputs of the network, $\hat \Bigs$ and $\hat \Bign$, are computed from $\Bigy$. Stacking the real and imaginary components $\Bigy_{\text{stack}} = [\Bigy_r, \Bigy_i]$ where subscript $r$ and $i$ denote real and imaginary parts, respectively, the mask estimation network aims to estimate a mask $\Bigmm_{\text{stack}} = [\Bigmm_r, \Bigmm_i] \in [\R^{F \times T \times C}, \R^{F \times T \times C} ]$. The complex multiplication between $\Bigy$ and $\Bigmm$ produces estimated speech $\hat \Bigs \in \mathbb{C}^{F \times T \times C}$~\cite{TFmask, maskchak}. Hence, $\Bigmm$ can be considered as the CRM for speech. Given $\Bigmm_{\text{stack}}$, the noise mask $\Bigmm^{\text{noise}}_{\text{stack}}$ is computed for the estimate of the noise $\hat \Bign$ as follows:
\begin{equation}\label{eq:gen}
\begin{aligned}
\Bigmm_r^{\text{noise}} &= 1 - \Bigmm_r,\qquad\quad \Bigmm_i^{\text{noise}} = -\Bigmm_i.
\end{aligned}
\vspace{-0.5em}
\end{equation}
The clean speech $\hat \Bigs$ and noise $\hat \Bign$ are estimated by the element-wise complex multiplication (denoted as $\ast$)
\begin{equation}\label{eq:gen}
\begin{aligned}
\hat \Bigs &= \Bigy * \Bigmm,\qquad\quad \hat \Bign = \Bigy * \Bigmm^{\text{noise}}.
\end{aligned}
\vspace{-0.5em}
\end{equation}

To train the network, we minimize the weighted $\ell_1$ loss of the audio in time domain and the magnitude of its spectrogram as follows:
\begin{equation}\label{eq:loss} 
\mathcal{L}(\smlu, \hat \smlu) = \sum_{u \in \{\s, \n\}} \alpha \| \smlu -  \hat \smlu \|_1 + \big\lVert |\Bigu| - |\hat \Bigu | \big\rVert_1,
\vspace{-0.5em}
\end{equation}
where $\alpha$ is determined based on the relative importance of the two error terms. The framework is trained in a \textit{supervised} manner, requiring ground truth speech and noise signals \cite{wang2018supervised}.
\vspace{-1mm}
\subsection{Network Architecture}

The mask estimation network is a variant of U-Net that consists of a series of blocks. The first block is a single unit of CA to perform beamforming-like operation on the input mixture, and the last block is a convolutional layer with ReLU non-linearity to generate the mask. We explain the middle blocks for the rest of this section.

We note that the real version of Channel-Attention Dense U-Net takes the magnitude of the STFT as its input, and estimates a real mask~\cite{TFmask}. This implies that the denoising of the noisy mixture is performed only with respect to the magnitude, hence the estimated clean speech contains phase of the noisy mixture.
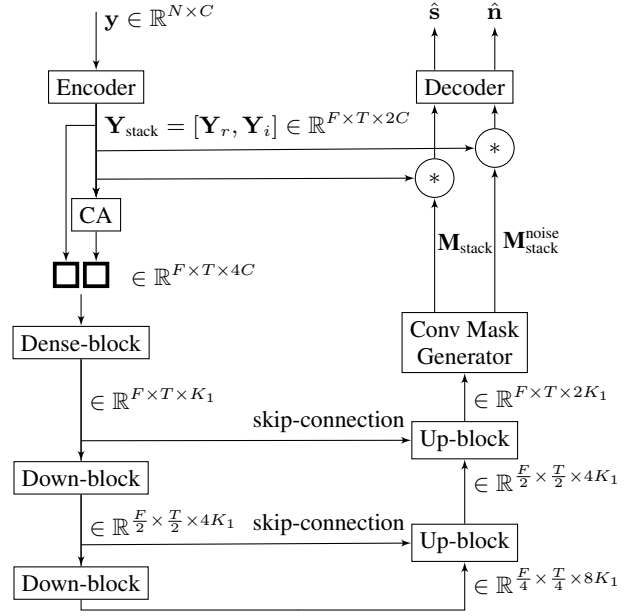
\begin{figure}[htb]
	\vspace*{-2mm}
	\begin{minipage}[b]{1.0\linewidth}
		\centering
		\tikzstyle{block} = [draw, fill=none, rectangle, 
		minimum height=1em, minimum width=1em]
		\tikzstyle{sum} = [draw, fill=none, circle, node distance=1cm]
		\tikzstyle{cir} = [draw, fill=none, circle, line width=1mm, minimum width=0.5cm, node distance=1cm]
		\tikzstyle{state} = [draw, fill=none, color=black, rectangle, line width=0.5mm, minimum width=0.7cm, minimum height=0.3cm, node distance=1cm]
		\tikzstyle{input} = [coordinate]
		\tikzstyle{output} = [coordinate]
		\tikzstyle{pinstyle} = [pin edge={to-,thin,black}]
		\begin{tikzpicture}[auto, node distance=2cm,>=latex']
		cloud/.style={
			draw=red,
			thick,
			ellipse,
			fill=none,
			minimum height=1em}
		\node [input, name=input] {};
		
		\node [block, fill=none, node distance=1cm, below of=input] (encoder) {$\text{Encoder}$};
		\node [block, fill=none, node distance=1.7cm, below of=encoder] (sa) {$\text{CA}$};

		\node [state, fill=none, minimum width=0.3cm, node distance=0.8cm, below of=sa] (x1a) {};
		\node [state, fill=none, minimum width=0.3cm, node distance=0.4cm, left of=x1a] (x1b) {};
		\node [rectangle, fill=none, minimum height=0.5cm, minimum width=0.3cm, node distance=0.2cm, left of=x1a] (x1m) {};
		\node [rectangle, fill=none, minimum height=0.5cm, minimum width=0.3cm, node distance=1.3cm, right of=x1a] (x1) {$\in \R^{F \times T \times 4C}$};
		
		\node [block, fill=none, node distance=0.9cm, below of=x1m] (dense) {$\text{Dense-block}$};
		\node [block, fill=none, node distance=1.8cm, below of=dense] (d1) {$\text{Down-block}$};
				
		\node [block, fill=none, node distance=1.4cm, below of=d1] (d2) {$\text{Down-block}$};
		
		\node [block, fill=none, node distance=5.1cm, right of=d2, above=8pt] (u2) {$\text{Up-block}$};
		\node [rectangle, fill=none, node distance=5.3cm, right of=d2, above=1pt] (u2right) {};
		
		\node [block, fill=none, node distance=5.1cm, right of=d1, above=7pt] (u1) {$\text{Up-block}$};

		\node [block, fill=none, node distance=5.1cm, right of=dense, align=center] (mask) {Conv Mask\\ Generator};
		\node [rectangle, fill=none, node distance=0.4cm, right of=mask, above=5pt] (maskright) {};
		\node [rectangle, fill=none, node distance=0.4cm, left of=mask, above=5pt] (maskleft) {};

		\node [sum, fill=none, minimum width=0.4cm, node distance=2.3cm, above of=maskright] (prodright) {$*$};
		\node [sum, fill=none, minimum width=0.4cm, node distance=1.90cm, above of=maskleft] (prodleft) {$*$};

		\node [block, fill=none, minimum width=0.4cm, node distance=4.9cm, right of=encoder] (decoder) {$\text{Decoder}$};
		\node [rectangle, fill=none, node distance=0.4cm, right of=decoder, below=0.1pt] (dright) {};
		\node [rectangle, fill=none, node distance=0.4cm, left of=decoder, below=0.1pt] (dleft) {};
		\node [rectangle, fill=none, node distance=0.4cm, right of=decoder, above=0.1pt] (dright_1) {};
		\node [rectangle, fill=none, node distance=0.4cm, left of=decoder, above=0.1pt] (dleft_1) {};
		
		\node [rectangle, fill=none, node distance=0.8cm, above of=dleft_1] (outy) {};
		\node [rectangle, fill=none, node distance=0.8cm, above of=dright_1] (outn) {};
		
		\node [rectangle, fill=none, node distance=0.1cm, above of=outy] (s) {$\hat \s $};
		\node [rectangle, fill=none, node distance=0.1cm, above of=outn] (n) {$\hat \n$};

		\draw [->] (input) -- node[pos=0.1] {$\y \in \R^{N \times C}$} (encoder);
		
		\draw [->] (encoder) -- node[name=e, pos=0.23, right] {$\Bigy_{\text{stack}} = [\Bigy_r, \Bigy_i] \in \R^{F \times T \times 2C}$} (sa);
		\draw [->] (encoder) -- node[name=ymask, pos=0.8, left] {} (sa);
		\draw [->] (encoder) -- node[name=nmask, pos=0.5, left] {} (sa);
		
		\draw [->] (sa) -- node[] {} (x1a);
		\draw [->] (e) -| node[] {} (x1b);
		
		\draw [->] (x1m) -- node[] {} (dense);
		\draw [->] (dense) -- node[pos=0.8, above=0.55cm, right] {$\in \R^{F \times T \times K_1}$} (d1);
		\draw [] (dense) -- node[name=densetod1, pos=0.8, left] {} (d1);
		
		\draw [->] (d1) -- node[pos=0.4, right] {$\in \R^{\frac{F}{2} \times \frac{T}{2} \times 4K_1}$} (d2);
		\draw [] (d1) -- node[name=d1tod2, pos=0.7, left] {} (d2);
		
		\node [rectangle, fill=none, node distance=3cm, right of=d2, below=7pt] (middle1) {};
		\node [rectangle, fill=none, node distance=2.77cm, right of=d2, below=7pt] (middle2) {};
		\draw[-] (d2) |- node[] {} (middle1);
		\draw[->] (middle2) -| node[pos=0.8, right] {$\in \R^{\frac{F}{4} \times \frac{T}{4} \times 8K_1}$} (u2);
		
		\draw [->] (d1tod2) -- node[pos=0.75] {skip-connection} (u2);
		\draw [->] (u2) -- node[pos=0.7, right] {$\in \R^{\frac{F}{2} \times \frac{T}{2} \times 4K_1}$} (u1);
		\draw [->] (densetod1) -- node[pos=0.75] {skip-connection} (u1);
		
		\draw [->] (u1) -- node[right] {$\in \R^{F \times T \times 2K_1}$} (mask);
		\draw [->] (maskright) -- node[right] {$\textbf{M}^{\text{noise}}_{\text{stack}}$} (prodright);
		\draw [->] (nmask) -- node[] {} (prodright);
	
		\draw [->] (maskleft) -- node[pos=0.6, right=-2pt] {$\textbf{M}_{\text{stack}}$} (prodleft);
		\draw [->] (ymask) -- node[] {} (prodleft);
		
		\draw [->] (prodright) -- node[] {} (dright);
		
		\draw [->] (prodleft) -- node[] {} (dleft);
		
		\draw [->] (dleft_1) -- node[] {} (outy);
		\draw [->] (dright_1) -- node[] {} (outn);

		\end{tikzpicture}
	\end{minipage}
	\vspace*{-6mm}
	\caption{Architecture of Channel-Attention Dense U-Net for $L=2$.}
	\label{fig:model}
	\vspace*{-7mm}
\end{figure}
\subsubsection{U-Net with DenseNet Blocks}
\label{sec:unet}
U-Net, a convolutional network previously proposed for image segmentation, is a popular network for source separation~\cite{waveUnet} and speech enhancement~\cite{phaseComplexUnet}. U-Net consists of a series of blocks ($L$ down-blocks and $L$ up-blocks), and skip-connections between down and up-blocks. Each down-block consists of a pooling layer for down-sampling, a convolutional layer, and an exponential non-linearity. Each up-block consists of an up-sampling through a transposed convolution with stride 2, a transposed convolutional layer, and an exponential non-linearity. 

\begin{figure}[htb]
	\vspace*{-2mm}
	\begin{minipage}[b]{1.0\linewidth}
		\centering
		\tikzstyle{block} = [draw, fill=none, rectangle, 
		minimum height=1em, minimum width=1em]
		\tikzstyle{sum} = [draw, fill=none, circle, node distance=1cm]
		\tikzstyle{cir} = [draw, fill=none, circle, line width=1mm, minimum width=0.5cm, node distance=1cm]
		\tikzstyle{state} = [draw, fill=none, color=black, rectangle, line width=0.5mm, minimum width=0.7cm, minimum height=0.3cm, node distance=1cm]
		\tikzstyle{input} = [coordinate]
		\tikzstyle{output} = [coordinate]
		\tikzstyle{pinstyle} = [pin edge={to-,thin,black}]
		\begin{tikzpicture}[auto, node distance=2cm,>=latex']
		cloud/.style={
			draw=red,
			thick,
			ellipse,
			fill=none,
			minimum height=1em}
		\node [input, name=input] {};
		\node [block, fill=none, node distance=0.7cm, below of=input] (pool) {$\text{Pooling}$};
		\node [block, fill=none, node distance=1.cm, below of=pool] (dense) {$\text{Dense-block}$};
		
		\node [block, fill=none, node distance=1.cm, below of=dense] (sa) {$\text{CA}$};
		
		\node [state, fill=none, minimum width=0.3cm, node distance=0.8cm, below of=sa] (x1a) {};
		\node [state, fill=none, minimum width=0.3cm, node distance=0.4cm, left of=x1a] (x1b) {};
		\node [rectangle, fill=none, minimum height=0.5cm, minimum width=0.3cm, node distance=0.2cm, left of=x1a] (x1m) {};
		\node [rectangle, fill=none, minimum height=0.5cm, minimum width=0.3cm, node distance=0.1cm, above of=x1m] (x1) {};
		
		\node [output, node distance=0.7cm, below of=x1m] (out) {};
		
		\draw [->] (input) -- node[pos=0.4, right=0pt] {$\in \R^{F \times T \times K}$} (pool);
		\draw [->] (pool) -- node[pos=0.5, right=0pt] {$\in \R^{\frac{F}{2} \times \frac{T}{2} \times K}$} (dense);
		\draw [] (dense) -- node[name=e, pos=0.5, right=0pt] {$\in \R^{\frac{F}{2} \times \frac{T}{2} \times 2K}$} (sa);
		\draw [->] (dense) -- node[name=e, pos=0.5, ] {} (sa);

		\draw [->] (sa) -- node[name=ymask, pos=0.8, left] {} (x1a);
		\draw [->] (e) -| node[name=nmask, pos=0.5, left] {} (x1b);
		\draw [] (x1m) -- node[right=5pt] {$\in \R^{\frac{F}{2} \times \frac{T}{2} \times 4K}$} (out);
		\draw [->] (x1) -- node[right=5pt] {} (out);

		\node [input, name=input, node distance=5cm, right of =x1b, below=45pt] {};
		
		\node [block, fill=none, node distance=0.7cm, above of=input] (up) {$\text{Up-sampling}$};
		\node [block, fill=none, node distance=1cm, above of=up] (conv) {$\text{Conv}$};

		\node [state, fill=none, minimum width=0.3cm, node distance=1cm, above of=conv] (x2a) {};
		\node [state, fill=none, minimum width=0.3cm, node distance=0.4cm, left of=x2a] (x2b) {};
		\node [rectangle, fill=none, minimum height=0.5cm, minimum width=0.3cm, node distance=0.2cm, left of=x2a] (x2m) {};
		\node [rectangle, fill=none, minimum height=0.5cm, minimum width=0.3cm, node distance=0.1cm, below of=x2m] (x2) {};	
	
		\node [block, fill=none, node distance=1.1cm, above of=x2] (dense) {$\text{Dense-block}$};
		\node [block, fill=none, node distance=1.cm, above of=dense] (sa) {$\text{CA}$};
				
		\node [state, fill=none, minimum width=0.3cm, node distance=0.8cm, above of=sa] (x1a) {};
		\node [state, fill=none, minimum width=0.3cm, node distance=0.4cm, left of=x1a] (x1b) {};
		\node [rectangle, fill=none, minimum height=0.5cm, minimum width=0.3cm, node distance=0.2cm, left of=x1a] (x1m) {};
		\node [rectangle, fill=none, minimum height=0.5cm, minimum width=0.3cm, node distance=0.1cm, below of=x1m] (x1) {};

		\node [output, node distance=0.7cm, above of=x1m] (out) {};
			
		\draw [->] (input) -- node[pos=0.4, right=0pt] {$\in \R^{\frac{F}{2} \times \frac{T}{2} \times 4K}$} (up);	
		\draw [->] (up) -- node[pos=0.4, right=0pt] {$\in \R^{F \times T \times 4K}$} (conv);
		\draw [->] (conv) -- node[name=skip, pos=0.4, right=0pt] {$\in \R^{F \times T \times K}$} (x2a);
		
		\draw [->] (x2) -- node[name=skip, pos=0.4, right=0pt] {$\in \R^{F \times T \times 2K}$} (dense);

		\draw [] (dense) -- node[name=e, pos=0.5, right=0pt] {$\in \R^{F \times T \times K}$} (sa);

		\node[input, node distance=1cm, left of=x2b] (down_skip) {};
		\draw [->] (down_skip) -- node[left=14pt] {skip} (x2b);
		
		\draw [->] (sa) -- node[name=ymask, pos=0.8, left] {} (x1a);
		\draw [->] (e) -| node[name=nmask, pos=0.5, left] {} (x1b);
		\draw [] (x1m) -- node[right=5pt] {$\in \R^{F \times T \times 2K}$} (out);
		\draw [->] (x1) -- node[right=5pt] {} (out);
		
		
		\node [rectangle, fill=none, node distance=0.3cm, above of=out] (b) {(b) Up-block};
		\node [rectangle, fill=none, node distance=4.4cm, left of=b] (a) {(a) Down-block};

		\end{tikzpicture}
	\end{minipage}
	\vspace*{-7mm}
	\caption{(a) Down Block, (b) Up Block. The skip connection is the output from the corresponding down-block.}
	\label{fig:downup}
	\vspace*{-5mm}
\end{figure}
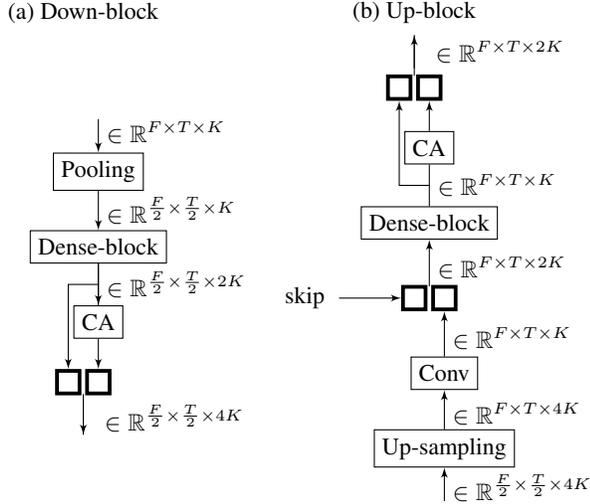

In Channel-Attention Dense U-Net, each convolutional layer in each block is replaced by a DenseNet block followed by a CA unit. The output of each down-block or up-block is the concatenation of the input and output of its CA unit. DenseNet applies convolution to the concatenation of several previous-layer feature maps, which eases the gradient flow in deep networks and helps each layer learn features that are not similar to the neighbouring layers~\cite{takahashi2017multi}. Figure~\ref{fig:model} shows the architecture of Channel-Attention Dense U-Net for when $L=2$, and Figure~\ref{fig:downup} shows the detail of down and up blocks.

\vspace{-2mm}
\subsubsection{Channel-Attention}
\label{sec:sa}

In this section, we introduce the Channel-Attention unit inspired by self-attention and beamforming. Self-attention is a mechanism for capturing global dependencies, which has gained attention in various fields such as machine translation~\cite{bahdanau2014neural}, natural language processing~\cite{parikhetal2016decomposable}, and image processing~\cite{saGANs}. 

Given input $\x \in \R^{\widetilde{F} \times \widetilde{T} \times 2\widetilde{C}}$, which is also the output of the DenseNet in mid blocks, our proposed CA unit transforms $\x$ into {\it key} $\kk(\x) \in \R^{\widetilde{F} \times d \times 2\widetilde{C}}$, {\it query} $\q(\x) \in \R^{\widetilde{F} \times d \times 2\widetilde{C}}$, and {\it value} $\smlv(\x) \in \R^{\widetilde{F} \times \widetilde{T} \times 2\widetilde{C}}$ feature maps through a convolution followed by an exponential non-linearity. Note that we use $\widetilde{(\cdot)}$ to indicate that due to down/up-sampling layer in each block, the dimensions of the input are different for CA unit at different blocks. The $\kk$, $\q$, and $\smlv$ are $1\times 1$ convolutional operators in the 2-dimensional space of $\widetilde{F} \times 2\widetilde{C}$ with $\widetilde{T}$ input channels and $d$, $d$, and $\widetilde{T}$ output channels, respectively, followed by an exponential non-linearity.

\begin{figure}[htb]
	\vspace*{-3mm}
	\begin{minipage}[b]{1.0\linewidth}
		\centering
		\tikzstyle{block} = [draw, fill=none, rectangle, 
		minimum height=2em, minimum width=2em]
		\tikzstyle{sum} = [draw, fill=none, circle, node distance=1cm]
		\tikzstyle{cir} = [draw, fill=none, circle, line width=1mm, minimum width=0.7cm, node distance=1cm]
		\tikzstyle{loss} = [draw, fill=none, color=black, ellipse, line width=0.5mm, minimum width=0.7cm, node distance=1cm]
		\tikzstyle{blueloss} = [draw, fill=none, color=black, ellipse, line width=0.5mm, minimum width=2.5cm, node distance=1cm, color=black]
		\tikzstyle{input} = [coordinate]
		\tikzstyle{output} = [coordinate]
		\tikzstyle{pinstyle} = [pin edge={to-,thin,black}]
		\begin{tikzpicture}[auto, node distance=2cm,>=latex']
		cloud/.style={
			draw=red,
			thick,
			ellipse,
			fill=none,
			minimum height=1em}
		\node [input, name=input] {};
		\node [rectangle, fill=none, node distance=0.001cm, right of=input] (stack) {$\x$};
		\node [rectangle, fill=none, node distance=1.1cm, above of=input] (up) {};
		\node [rectangle, fill=none, node distance=0.55cm, above of=input] (middle) {};
		\node [rectangle, fill=none, node distance=0.0001cm, right of=input] (down) {};
		\node [rectangle, fill=none, node distance=0.9cm, below of=input] (dd) {};
		
		\node [block, node distance=1.2cm, right of=down] (query) {$\q(\x)$};
		\node [rectangle, node distance=2.4cm, right of=down, above=0.001cm] (queryname) {{\it query}};
		
		\node [block, node distance=1.2cm, right of=up] (key) {$\kk(\x)$};
		\node [rectangle, node distance=2.4cm, right of=up, above=0.001cm] (keyname) {{\it key}};
		
		\node [block, node distance=1.2cm, right of=dd] (value) {$\smlv(\x)$};

		\node [sum, right of=middle, node distance=3.2cm] (p1) {$\times$};
		\node [rectangle, node distance=3.8cm, right of=middle, above=0.001cm] (for) {$\Bigp$};
		\node [rectangle, node distance=5.9cm, right of=middle, above=0.001cm] (sa) {$\ww$};
		\node [rectangle, node distance=1.45cm, below of=sa] (valuename) {{\it value}};

		\node [block, right of=p1, node distance=1.6cm] (soft) {softmax};
		\node [sum, right of=query, node distance=5.2cm] (p2) {$\times$};
		
		\node [rectangle, fill=none, right of=p2, node distance=0.6cm, above=1pt] (outname) {$\mathbf{o}$};
		\node [rectangle, fill=none, right of=p2, node distance=1.cm] (out) {};
		
		\draw [->] (stack) |- node[] {} (key);
		\draw [->] (stack) -- node[] {} (query);
		\draw [->] (stack) |- node[] {} (value);
		
		\draw [->] (key) -| node[] {} (p1);
		\draw [->] (query) -| node[] {} (p1);
		
		\draw [->] (p1) -- node[] {} (soft);
		\draw [->] (soft) -| node[] {} (p2);
		\draw [->] (value) -| node[] {} (p2);
		
    		\draw [->] (p2) -- node[] {} (out);

		\end{tikzpicture}
	\end{minipage}
	\vspace*{-6mm}
	\caption{CA unit. Given input $\x$, CA computes the attention mask $\ww$ and apply it to {\it value}, a variant of the input.}
	\label{fig:sa}
	\vspace*{-5mm}
\end{figure}
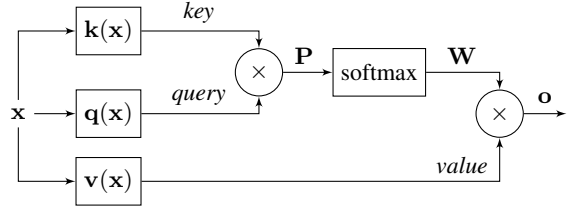

We treat the {\it key}, {\it query}, and {\it value} as stack of real and imaginary where the first $\widetilde{C}$ channels are real and the second $\widetilde{C}$ channels are imaginary. For a given frequency bin $f$, we define the {\it key} and {\it query} as $\kk_f(\x) \in \mathbb{C}^{d \times \widetilde{C}}$ and $\q_f(\x) \in \mathbb{C}^{d \times \widetilde{C}}$. The CA mechanism computes the similarity matrix, $\Bigp = [\Bigp_1, \ldots, \Bigp_{\widetilde{F}}] \in \mathbb{C} ^{\widetilde{F} \times \widetilde{C} \times\widetilde{C}}$ between {\it key} and {\it query} for every frequency bin as follows:

\begin{equation}\label{eq:fg}
\Bigp_f = \kk_f(\x)^{\text{T}} \q_f(\x) \in \mathbb{C}^{\widetilde{C} \times \widetilde{C}},\ \textrm{for } f = 1,\ldots, \widetilde{F}.\\
\end{equation}
Having $\kk_f(\x) = [\kk_{f1},\ldots, \kk_{f\widetilde{C}}] $ and $\q_f(\x) = [\q_{f1},\ldots, \q_{f\widetilde{C}}] $,  for $f = 1,\ldots, \widetilde{F}$, the similarity matrix is,
\begin{equation}\label{eq:corr}
\Bigp_f = \kk_f^{\text{T}} \q_f =
  \begin{bmatrix}
    \kk_{f1}^{\text{T}} \q_{f1} & \ldots & \kk_{f1}^{\text{T}} \q_{f\widetilde{C}} \\
    \kk_{f2}^{\text{T}} \q_{f1} & \ldots & \kk_{f2}^{\text{T}} \q_{f\widetilde{C}} \\
     \vdots & \ldots &  \vdots \\
     \kk_{f\widetilde{C}}^{\text{T}} \q_{f1} & \ldots & \kk_{f\widetilde{C}}^{\text{T}} \q_{f\widetilde{C}} \\
  \end{bmatrix}.\\
\end{equation}
The attention weights matrix $\ww \in \mathbb{C}^{\widetilde{F} \times \widetilde{C} \times \widetilde{C}}$ is normalized (by softmax function) $\Bigp$ with respect to the second dimension. The weight matrix entry is thus given as 
\begin{equation}\label{eq:attention}
|w_{f,c,c^\prime}| = \frac{\e^{|p_{f,c,c^\prime}|}}{\sum_{c=1}^{\widetilde{C}} \e^{|p_{f,c,c^\prime}|}},\qquad
\angle w_{f,c,c^\prime} =  \angle p_{f,c,c^\prime},
\end{equation}
$\textrm{for } f = 1,\ldots, F, \textrm{and } c,c^{\prime} = 1,\ldots, \widetilde{C}$. The output of the attention unit for frequency $f$ is the concatenation of the real and imaginary parts of $\textbf{o}_f$ computed as follows:
\begin{equation}\label{eq:out_sa} 
\textbf{o}_f = \smlv_f(\x) \ww_f \in \mathbb{C}^{\widetilde{T} \times \widetilde{C}},
\end{equation}
where $\smlv_f(\x) \in  \mathbb{C}^{\widetilde{T} \times \widetilde{C}}$, and $\ww_f \in \mathbb{C}^{\widetilde{C} \times \widetilde{C}}$. For real-valued input, multiplication and similarity operations happen in real domain. Figure~\ref{fig:sa} shows the detailed architecture of CA. 

\subsection{Connection of Channel-Attention to Beamforming}

The motivation for incorporating the CA concept into our framework is two-fold. First, inspired by the traditional beamformers which linearly combine multichannel mixtures to produce a clean signal estimate, we expect the trained CA unit to learn to `optimally' combine multichannel information to produce a clean speech signal. Specifically, the fact that a CA unit is applied to features maps at every layer, and that nonlinearity layers exist throughout the architecture suggests that this combination is not confined to the linear regime.


In Eq.~(\ref{eq:corr}), each column $c$ resembles beamforming weights as if channel $c$ is chosen as reference. Therefore, in Eq.~(\ref{eq:out_sa}), $\smlv_f(\x)$ can be seen as a variant of the input signal to CA, and $\ww_f$ decides which channel of $\smlv_f(\x)$ to pay more attention to. Indeed, our proposed CA can be seen as a mechanism to automatically pick a reference channel and perform beamforming. Interestingly, we observe that the attention weights in a trained model learn to represent the signal-to-noise-ratio (SNR) ({\it importance}) of each feature map.


We verified this behaviour by examining the weights of the trained CA unit $\ww$, located right after the encoder, from the trained CA unit of real Channel-Attention Dense U-Net for the following two input scenarios: 1) a noisy mixture from the CHiME-3 dataset~\cite{chime} and 2) a toy example with the simulated input where channel 1 has the highest SNR among all channels. We chose to examine the real network instead of the complex network, for easier interpretation and visualization.


\begin{figure}[!htb]
	\vspace*{-2mm}
	\begin{minipage}[b]{1.0\linewidth}
		\centering
		\tikzstyle{block} = [draw, fill=none, rectangle, 
		minimum height=2em, minimum width=2em]
		\tikzstyle{sum} = [draw, fill=none, circle, node distance=1cm]
		\tikzstyle{cir} = [draw, fill=none, circle, line width=1mm, minimum width=0.7cm, node distance=1cm]
		\tikzstyle{loss} = [draw, fill=none, color=black, ellipse, line width=0.5mm, minimum width=0.7cm, node distance=1cm]
		\tikzstyle{blueloss} = [draw, fill=none, color=black, ellipse, line width=0.5mm, minimum width=2.5cm, node distance=1cm, color=black]
		\tikzstyle{input} = [coordinate]
		\tikzstyle{output} = [coordinate]
		\tikzstyle{pinstyle} = [pin edge={to-,thin,black}]
		\begin{tikzpicture}[auto, node distance=2cm,>=latex']
		cloud/.style={
			draw=red,
			thick,
			ellipse,
			fill=none,
			minimum height=1em}
		\node [input, name=input] {};
		
		\node [rectangle, fill=none, node distance=3cm, right of=input] (A) {$\includegraphics[width=0.8\linewidth]{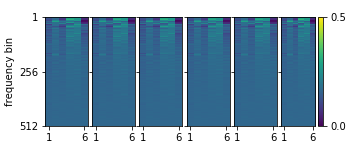}$};
		
		\node [rectangle, fill=none, node distance=2.5cm, below of=A] (C) {$\includegraphics[width=0.8\linewidth]{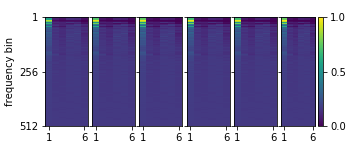}$};
		
		\node [rectangle, fill=none,  node distance=4.2cm,  left of=A] (text) {(a)};
		
		\node [rectangle, fill=none,  node distance=4.2cm,  left of=C] (text) {(b)};
		
		\end{tikzpicture}
	\end{minipage}
	\vspace*{-8mm}
	\caption{(a)  $\ww$ from the CHiME-3 dataset (b) $\ww$ from the SNR example.}
	\label{fig:snr}
	\vspace*{-2mm}
\end{figure}
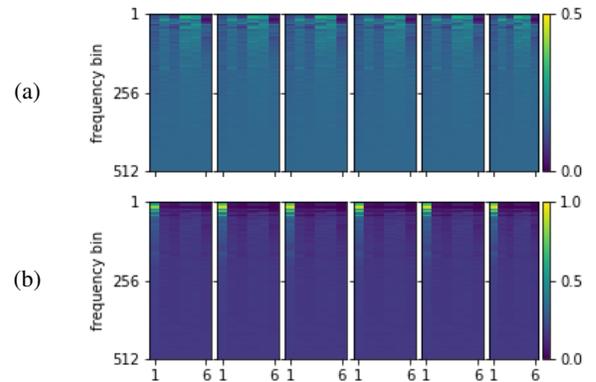
Figure~\ref{fig:snr}(a) shows the $[\ww_{F \times C \times 1}, \ldots, \ww_{F \times C \times C}]$ for the CHiME-3 data example, where we observe that channel $5$ has received the most attention. This matches with the fact that channel $4$ and $5$ of CHiME-3 recordings have the highest SNR on average. Another interesting observation is that CA learns to pay more attention to low frequencies, which are known to contain the majority of the speech information.

Figure~\ref{fig:snr}(b) demonstrates the similar results for a toy example. In this case, we clearly observe that channel $1$, the channel with the highest SNR, gets the most attention.
\vspace{-2mm}
\section{Experiments}
\label{sec:exp}
\vspace{-2mm}
\subsection{Dataset}

We used the publicly available CHiME-3 dataset~\cite{chime}, made available as part of a speech separation and recognition challenge, for training and evaluating speech enhancement performance. The dataset is a $6$-channel ($C = 6$) microphone recording of talkers speaking in a noisy environment, sampled at $16$ kHz. It consists of $7{,}138$, $1{,}640$, and $1{,}320$ simulated utterances with an average length of ${3}$ s for training, development, and test, respectively. At every iteration of training, a random segment of length $N = 19{,}200$ is selected from the utterance, and a random attenuation of the background noise in the range of ${[-20, 0]}\text{ dB}$ is applied as a data augmentation scheme. This augmentation was done to make the network robust against various SNRs.
\vspace{-2mm}
\subsection{Training and Network Parameters}
The encoder and decoder are initialized with STFT and Inverse-STFT coefficients, respectively, using a Hanning window of length $1{,}024$, hence $F = 512$ (we discard the last bin, since, for the downsampling step of the network, the number of frequency bins needs to be even), and hop size of ${256}$. Consequently, the number of time frames for each input is $T = 80$. We design the network to have $4$ down-blocks and $4$ up-blocks $(L = 4)$ where the kernel size for all convolutions is $2 \times 2$. The number of convolutional filters in the first layer is set to ${32}$, with a maximum of $256$ possible number of filters at every convolution. For all the CA units inside the network, we set the depth of {\it query} and {\it key} to $d = {20}$.

The network is trained with ADAM optimizer with learning rate of ${10^{-4}}$, and batch size of $8$. For the loss function (Eq.~(\ref{eq:loss})), we set $\alpha$ such that the error in time domain, $\| \smlu -  \hat \smlu \|_1$, is twice as important as the error in the magnitude of the spectrogram, $\big\lVert |\Bigu| - |\hat \Bigu | \big\rVert_1$. This is done based on loss magnitude at the beginning of training.

 

\vspace{-2mm}
\subsection{Evaluation}


We evaluated the network performance with the following metrics: signal-to-distortion ratio (SDR) using BSS Eval library~\cite{bsseval} and Perceptual Evaluation of Speech Quality (PESQ) - more specifically the wideband version recommended in ITU-T P.862.2 (–0.5 to 4.5).

Given the source estimates $\hat \s$ from all $C$ channels at the output, we computed the posterior SNR for each channel and selected the channel with the highest posterior SNR as the final estimate.

\vspace{-0.5em}
\section{Results}
\label{sec:res}
\vspace{-0.5em}
We trained four networks as follows:

\begin{itemize}
	\vspace{-0.5em}
\item \textbf{U-Net (Real)}: U-Net without any dense blocks, which performs magnitude ratio masking.
\item \textbf{Dense U-Net (Real)}: U-Net (Real) with dense blocks ($D$$=$$4$).
\item \textbf{Dense U-Net (Complex)}: U-Net with dense blocks ($D$$=$$4$), which takes real and imaginary part of STFT as input and performs complex ratio masking.
\item \textbf{CA Dense U-Net (Complex)}: Dense U-Net (Complex) with Channel-Attention.
\end{itemize}

\begin{table}[htb]
\setlength{\tabcolsep}{5pt}
\renewcommand{\arraystretch}{1.3}
\caption{Performance of trained networks on CHiME-3 dataset.}
\label{tab:res}
\centering
\begin{tabular}{c|c|c||c|c}
    \hline
    \hline
    \multirow{1}{*}{} &
      \multicolumn{2}{c}{sim-dev}  \vline &
      \multicolumn{2}{c}{sim-test} \\
     \hline
     Methods & SDR & PESQ & SDR & PESQ \\
     \hline
     Channel-5 (Noisy) &5.79 &1.27 &6.50 &1.27 \\
    \hline
    U-Net (Real) & 14.651 & 2.105 & 15.967 & 2.176 \\
    \hline
    Dense U-Net (Real) & 14.901 & 2.242 & 16.855 &  2.378 \\
    \hline
   Dense U-Net (Complex)& 16.962 & 2.33 & 18.402 &  2.404 \\
    \hline
   CA Dense U-Net (Complex) & {\bf 17.169} & {\bf 2.368} &  {\bf 18.635} & {\bf 2.436}
\end{tabular}
\vspace{-6mm}
\end{table}
\begin{table}[htb]
\setlength{\tabcolsep}{5pt}
\renewcommand{\arraystretch}{1.3}
\caption{Performance comparison of Channel-Attention Dense U-Net with state-of-the-art results on CHiME 3.}
\label{tab:res_compare}
\centering
\begin{tabular}{c|c|c|c|c}
    \hline
    \hline
    \multirow{1}{*}{} &
      \multicolumn{2}{c}{sim-dev}  \vline &
      \multicolumn{2}{c}{sim-test} \\
	\hline
    Methods & SDR & $\Delta$PESQ & SDR & $\Delta$PESQ \\
    \hline
    NMF B~\cite{nmfb} & - & - & 16.16 & 0.52 \\
    \hline
    Forgetting F~\cite{forget} & 16.07 & - & - & -\\
    \hline
    Neural B~\cite{erdogan} & 15.80 & 0.92 & 15.12 & 1.02\\
    \hline
    CA Dense U-Net & {\bf 17.169} & {\bf 1.09} & {\bf18.635} & {\bf 1.16} \\
\end{tabular}
\vspace{-4mm}
\end{table}
 
Table~\ref{tab:res} demonstrates the improvement in the performance of the network, as we add a new component to the architecture, such as dense-blocks, complex ratio masking scheme, and finally, the Channel-Attention. We note that for U-Net (Real), and Dense U-Net (Real), the only spatial information network has access to is ILD, the level difference between the channel. Hence the performance improvement from Dense U-Net (Real) to Dense U-Net (Complex) is primarily for two reasons: a) access to IPD information, and b) complex ratio masking instead of magnitude ratio masking. Finally, we observe that Channel-Attention improves the performance of Dense U-Net (Complex) further.






We compare the performance of our method to the following three state-of-the-art methods on CHiME-3 dataset:

\begin{itemize}
\item \textbf{Neural Beamforming}~\cite{erdogan}: An MVDR beamforming with mask estimation through bidirectional-LSTM.
\item \textbf{NMF-Informed Beamforming}~\cite{nmfb}: An online MVDR beamforming through the decomposition of TF bins of the mixture into the sum of speech and noise, by performing non-negative matrix factorization (NMF).
\item \textbf{Forgetting Factor Optimization}~\cite{forget}: An MVDR beamforming with simultaneous estimation of TF masks and forgetting factors.
\end{itemize}


Table~\ref{tab:res_compare} shows the results where $\Delta$PESQ represents PESQ improvement with respect to the channel 5 of the noisy mixtures (row 1 in Table~\ref{tab:res}). Results for the competing methods are taken from the corresponding papers and the missing entries in the table indicate that the metric is not reported in the reference paper. Overall, our proposed approach significantly outperforms state-of-the-art results on the CHiME-3 speech enhancement task.

\vspace{-2mm}
\section{Conclusion}
\label{sec:con}

This paper proposed a channel-attention mechanism inspired by beamforming for speech enhancement of multichannel recordings. The paper combined time-frequency masking~\cite{TFmask}, UNet~\cite{unet}, and DenseNet~\cite{denseNet} into a unified network along with channel-attention mechanism. Our interpretation of the channel-attention mechanism is that the network performs recursive \emph{non-linear} beamforming on the data represented in a latent space. We showed that the proposed network outperforms all the published state-of-the-art algorithms on the CHiME-3 dataset. 


\balance
\bibliographystyle{IEEEbib}
\bibliography{icassp20-aws}

\end{document}